\documentclass[aps]{revtex4}
\usepackage{graphicx}
\usepackage{color}
\usepackage[T1]{fontenc}
\usepackage[normalem]{ulem}

\RequirePackage{amsfonts,amssymb,amsmath}
\newtheorem{definição}{Definition}[section]
\newtheorem{teorema}{Theorem}[section]
\newtheorem{exemplo}{Example}[section]

\begin{document}
\title{  Quantum Singularities in Static Spacetimes  }
\author{Jo\~ao Paulo M. Pitelli} 
\email{e-mail:pitelli@ime.unicamp.br}
\author{Patricio S. Letelier} 
 \email{e-mail: letelier@ime.unicamp.br} 
\affiliation{
Departamento de Matem\'atica Aplicada-IMECC,
Universidade Estadual de Campinas,
13081-970 Campinas,  Sao Paulo, Brazil}

\begin{abstract}
We review the mathematical framework necessary to understand the physical content of quantum singularities in static spacetimes. We present many examples of classical singular spacetimes and study their singularities by using wave packets satisfying Klein-Gordon and Dirac equations. We show that in many cases the classical singularities are excluded when tested by quantum particles but unfortunately there are other cases where the singularities remain from the quantum mechanical point of view. When it is possible we also find, for spacetimes where quantum mechanics does not exclude the singularities, the boundary conditions necessary to turn the spatial portion of the wave operator into self-adjoint and emphasize their importance to the interpretation of quantum singularities. 
\end{abstract}

\maketitle
{\small
\hspace{1.6cm}{\it keywords:} quantum singularities, von Newmann theorem, self-adjointness}
\section{Introduction}

Classical singularities in general relativity are indicated by incomplete geodesics or incomplete paths of bounded acceleration \cite{hawking}. There are three types of singularities \cite{helliwell2,konkowski}: the quasi regular singularity, where no observer sees any physical quantities diverging even if its world line reaches the singularity (for example, the singularity in the spacetime of a cosmic string); the scalar curvature singularity, where every observer near the singularity sees physical quantities diverging (for example, the singularity in the Schwarzchild spacetime or the big bang singularity in FRW cosmology); the non scalar curvature singularity, where there are some curves in which the observers experience unbounded tidal forces (whimper cosmologies are a good example). In general relativity, singularities are not part of the spacetime since the spacetime is differentiable by definition, so the future of a test particle which reaches a singular point at a finite proper time is unpredictable unless some extra information is imposed. Moreover, in the classical singular points the laws of physics are no longer valid since they are formulated having a smooth classical spacetime background.

In order to try to avoid this conclusion it was conjectured the cosmic censorship hypothesis \cite{penrose}, which says basically that nature abhors naked singularities. Therefore, every singularity which appears as a solution of Einstein equation must be hidden by an event horizon. This hypothesis assures us that no observer at infinity or at least careful enough not to fall into a black hole will see any effect caused by the singularity. As Hawking said \cite{hawking2}, this hypothesis is selfish because it ignores the question of what happens to an observer who does fall through an event horizon and, despite the fact that it has not been proved, many authors suspect that in a fully classical context this conjecture is correct.  But recently, some examples of violation of the cosmic censorship hypothesis have been found, even when the spacetime is filled with reasonable matter, for example when gravity is coupled to a scalar field with potential $V(\phi)$ satisfying the positive energy theorem \cite{horowitz2}.

Under very reasonable conditions (the energy conditions) which state that gravity is always attractive, singularities are inevitable to general relativity. Since general relativity can not escape this burden, we hope that a complete theory of quantum gravity will overcome this situation, teaching us how to deal with such spacetime near the singularities or excluding the singularities at all. While such a theory does not exist, there are many attempts to incorporate quantum mechanics into general relativity. 

In this paper we use quantum field theory in curved spacetimes and, analogous to the classical case, we follow the ideas of Horowitz and Marolf \cite{horowitz} and say that a spacetime is quantum mechanically singular if the evolution of a wave packet representing an one particle state is not uniquely determined by the initial wave packet. In this case, the future of the quantum particle is also unpredictable unless we state some extra information, a boundary condition near the singularity to be more precise. Because of the general belief that the cosmic censorship is valid, we study here only naked singularities, i.e., singularities which are not hidden by an event horizon. A very interesting case where naked singularities appears is due to a topological defect, which are  characterized by a null curvature tensor everywhere, except on a submanifold, where it is proportional to a Dirac delta function. This case will also be studied here.

This paper is organized as follows: in Sec. 2
   we present a brief summary of the theory of unbounded operators in Hilbert spaces, giving emphasis to a theorem due to von Neumann which will provide a method to decide if the spacetime is quantum mechanically singular or regular.
 In Sec. 3
 we introduce the concept of quantum singularities. 
In Sec. 4
 we show several examples of the quantum behavior of test particles in classical singular spacetimes found in the literature. Finally, in 
Sec.  5
we discuss the examples presented in the previous section.

\section{Mathematical Framework}
A Hilbert space over the complex field $\mathbb{C}$ is a vector space provided with an inner product
\begin{equation}
\begin{aligned}
\left<\;,\,\right>:&H\times H &\longrightarrow &\,\mathbb{C}\\
                     &(\varphi,\psi) &\longmapsto &\left<\varphi, \psi \right>,
\end{aligned}
\end{equation}
which is anti linear in the first input and linear in the second one. Moreover, this space is complete in the norm $\left\|\cdot\right\|$, defined by the inner product by $\left\|\psi\right\|=\left<\psi,\psi\right>^{1/2}$, complete in the sense that every Cauchy sequence converges.

A linear operator in a Hilbert space $H$ is an operator satisfying the following property:
\begin{equation}
T(\alpha \psi+\varphi)=\alpha T\psi+T\varphi\qquad \forall \, \psi, \varphi \in H;\, \alpha \in \mathbb{C}.
\label{oplinear}
\end{equation}

Nevertheless, a crucial detail, which sometimes quantum mechanical textbooks do not mention, is that an operator is not currently defined if its domain is not specified. Operators with different domains must be considered distinct. The formal definition is: a linear operator $T:D(T)\rightarrow H$ is a linear map from a subset $D(T)\subseteq H$, which is called domain of $T$, into a subset $R(T)\subseteq H$, which is called range of $T$. The domain and the range of $T$ are vector subspaces of $H$ \cite{richtmyer}.

An operator  $T:D(T)\rightarrow H$  is said to be bounded if there exists a number $c>0$ such that $\forall\,\psi\in D(T)$ we have
\begin{equation}
\left\|T\psi\right\|\leq c\left\|\psi\right\|.
\end{equation}

By the representation theorem of Riesz-Fréchet \cite{richtmyer,kreyszig} we know that for a bounded operator $T:H\rightarrow H$, the Hilbert adjoint operator $T^{\ast}:H\rightarrow H$ defined by the equality
\begin{equation}
\left<T^{\ast}\varphi,\psi\right>=\left<\varphi,T\psi\right>\;\;\;\;\;\forall\,\psi,\varphi\in H
\end{equation}
exists and it is bounded.

However, it is a fact that many operators we work in quantum mechanics are unbounded, like the position and the momentum operators. For unbounded operators we do not have such a theorem like the Riesz-Fréchet theorem,  but we have  a weaker definition of Hilbert adjoint operator.
\begin{definição}[Hilbert adjoint operator]
Let $T:D(T)\rightarrow H$ be a linear operator densely defined in a Hilbert space $H$. The Hilbert adjoint operator is defined as follows: the domain $D(T^{\ast})$ of $T^{\ast}$ consists of all $\varphi\in H$ for which there is $\varphi^{\ast}\in H$ such that, for all $\psi\in D(T)$,
\begin{equation}
\left<\varphi^{\ast},\psi\right>=\left<\varphi,T\psi\right>.
\end{equation}

For each $\varphi\in D(T)$, the action of $T^{\ast}$ is given by
\begin{equation}
T^{\ast}\varphi=\varphi^{\ast}.
\end{equation}
\end{definição}

In the theorem above, an operator is said to be densely defined in the Hilbert space $H$ if $D(T)$ is dense in $H$ ($\overline{D(T)}=H$, where $\overline{D(T)}$ denotes the closure of $D(T)$). It is a necessary condition for $T^{\ast}$ to be a map, i.e.,  for $T^{\ast}$ associate a unique $\varphi^{\ast}$ to each $\varphi\in D(T^{\ast})$ \cite{richtmyer}.

An operator $T:D(T)\rightarrow H$ is called self adjoint if $T=T^{\ast}$. Note that the domains must be the same, i.e., $D(T)=D(T^{\ast})$. 

A self-adjoint operator is also symmetric, i.e., it satisfies
\begin{equation}
\left<\varphi,T\psi\right>=\left<T\varphi,\psi\right>
\end{equation}
$\forall\,\varphi,\psi\in D(T)$.

The concept of closed linear operators will be important in what follows.
\begin{definição}[Closed linear operator]
Let $T:D(T)\rightarrow H$ be a linear operator on a Hilbert space $H$. $T$ is said to be closed if
\begin{equation}
\left\{ \begin{array}{l}
\psi_{n}\rightarrow \psi \qquad [\psi_{n}\in D(T)]\\
T{\psi_{n}}\rightarrow \varphi
\end{array} \right.
\end{equation}
implies $\psi \in D(T)$ and $T\psi=\varphi$.
\end{definição}

An operator $T$ is {\bf closable} if it has a closed extension. Every closed operator has a smallest closed extension, called its {\bf closure}, which we denote by $\overline{T}$.

Let us now introduce the concept of continuous one-parameter unitary group. This will be the case for the evolution operator in quantum mechanics.
\begin{definição}
A one-parameter function operator $U(t)$ is called a continuous one-parameter unitary group if
\begin{itemize}
\item [a)] For each $t\in \mathbb{R}$, $U(t)$ is a unitary operator and $U(t+s)=U(t)U(s)\,\forall\,t,s\in\mathbb{R}$.
\item [b)] If $\psi\in H$ and $t\to t_0$ then $U(t)\psi\to U(t_0)\psi$.
\end{itemize}
\end{definição}

There is a theorem due to M. Stone which relates $U(t)$ with the exponential of a self-adjoint operator \cite{reed}.
\begin{teorema}[Stone's theorem]
Let $U(t)$ be a one parameter continuous unitary group in a Hilbert space $H$. Then exists a self-adjoint operator $A$ in $H$ such that $U(t)=e^{itA}$.
\end{teorema}

In non-relativistic quantum mechanics, the equation which governs the physical system is the Schr\"odinger equation
\begin{equation}
\mathcal{H}\left.|\psi\right>=-i\frac{\partial}{\partial t}\left.|\psi\right>,
\end{equation}
where $\mathcal{H}$ is the Hamiltonian operator of the system (suppose $\mathcal{H}$ is time independent).

If the operator $\mathcal{H}$ is self-adjoint, the state of the system at instant $t$ can be found by
\begin{equation}
\left.|\psi(t)\right>=U(t)\left.|\psi(0)\right>=e^{-i\mathcal{H}t/\hbar}\left.|\psi(0)\right>
\end{equation}

Physical motivations give rise to an expression to  the system Hamiltonian operator. Usually it is an operator with partial derivatives in an appropriate $L^2$ space. At first, the domain of the operator is not specified and it is simple to find a domain where the Hamiltonian operator is a well-defined symmetric operator (self-adjoint or not). Then, if the closure of $\overline{\mathcal{H}}$ is self-adjoint we can use it. Otherwise, if the closure of $\mathcal{H}$ is not self-adjoint we seek to figure out how many self-adjoint extensions does it has (possibly none).

To answer this question another two definitions will be necessary \cite{reed}.
\begin{definição}[Essentially self-adjoint operators]
A linear symmetric operator $T$ is called essentially self-adjoint if its closure $\overline{T}$ is self-adjoint.
\end{definição}

It can be shown that if $T$ is essentially self-adjoint it has a unique self-adjoint extension.
\begin{definição}
Let $T$ be a symmetric operator. Let
\begin{equation}
\begin{aligned}
&\mathcal{K}_+=\text{Ker}(i-T^{\ast})\\
&\mathcal{K}_-=\text{Ker}(i+T^{\ast}).
\end{aligned}
\end{equation}
$\mathcal{K}_+$ and $\mathcal{K}_{-}$ are called the {\bf deficiency subspaces} of $T$. The pair of numbers $n_+$,$n_-$, given by $n_+(T)=\text{dim}[\mathcal{K}_+]$, $n_-(T)=\text{dim}[\mathcal{K}_-]$ are called the {\bf deficiency indices} of $T$.
\end{definição}

The following theorem will give us a criterion to decide if the operator is essentially self-adjoint or not, and a method to find its self-adjoint extensions \cite{reed2}.
\begin{teorema}[Criterion for essentially self-adjoint operators]
Let $T$ be  a symmetric operator with deficiency indices $n_+$ and $n_-$. Then, 
\begin{itemize}
\item[(a)] If $n_+=n_-=0$, $T$ is essentially self-adjoint.
\item[(b)] If $n_+=n_-\geq 1$,  $T$ has an infinite number of self-adjoint extensions parametrized by a $n\times n$ matrix.
\item[(c)] If $n_+\neq n_-$,  $T$ does not have self-adjoint extensions.
\end{itemize}
\label{theorem2}
\end{teorema}

So, in order to determine if a symmetric operator $T$ is essentially self-adjoint we must solve the pair of equations
\begin{equation}
(T^{\ast}\mp i)\psi=0
\label{criterion}
\end{equation}
and count the number of independent solutions in $H$, i.e., the dimension of $\text{Ker}(T^{\ast}\mp i)$. Theorem $X.2$ of Ref. \cite{reed2} states that the self-adjoint extensions of the operator are represented by
 the one-parameter family of the extended domains of operator $T$ given by
\begin{equation}
D^{U}=\{\psi=\phi +\phi^{+}+U\phi^{-}:\phi \in D(T)\},
\label{neumann}
\end{equation}
where
\begin{equation}
T^{\ast}\phi^{\pm}=\pm i \phi^{\pm},
\end{equation}
$\phi^{\pm}\in H$, and $U$ is an isometry from $\ker(T^{\ast}-i)$ to 
 $\ker(T^{\ast}+i)$.  
 
As an example of everything we said before let us study the momentum operator given by
\begin{exemplo}[Momentum operator]
Let $H=L^2(0,1)$ and $T$ defined by
\begin{equation}\begin{aligned}
D(T)&=C_{0}^{\infty}(0,1)\\
Tf&=-if'.\label{momentum}
\end{aligned}\end{equation}

First of all, it is easy to check that $T$ is a symmetric operator since $\forall\,f,g\in D(T)$
\begin{equation}
\left<g,Tf\right>=\int_{0}^{1}{\overline{g}(-if')}=-i\underbrace{gf\arrowvert_{0}^{1}}_{=0}+i\int_{0}^{1}{\overline{g'}f}=\int_{0}^{1}{\overline{-ig'}f}=\left<Tg,f\right>,
\end{equation}
and, since $C_{0}^{\infty}(0,1)$ is dense in $L^{2}(0,1)$, $T$ is densely defined in $L^2(0,1)$.

In order to find the Hilbert adjoint operator we must look for all pairs $(g,g^{\ast})\in D(T^{\ast})\times L^{2}(0,1)$ such that $\forall\,f\in D(T)$
\begin{equation}
\left<g,Tf\right>=\left<g^{\ast},f\right>.
\end{equation}
So
\begin{equation}
\begin{aligned}
\left< g^{\ast}, f \right>&=\left< g,Tf \right>= \left< g,-if' \right>=-i \left< g,f' \right>=\\ &=-i\int_{a}^{b}{\overline{g}f'dx}=-i\overline{g}f\arrowvert^{1}_{0}+i\int_{a}^{b}{\overline{g'}f}=\int_{a}^{b}{\overline{-ig'}fdx}=\left< -ig',f\right>, 
\end{aligned}
\end{equation}
therefore $T^{\ast}g=-ig'$. But now, if $g_1\in L^2(0,1)$ and $g_1^{\ast}=-ig_1'\in L^2$, then (remember $f$ has compact support in $(0,1)$)
\begin{equation}
\left< g_{1}^{\ast},f \right>=\left<  -ig_{1}',f \right>=i \left< g_{1}', f \right>=-i \left< g_{1},f' \right> = \left< g_{1},-if'\right>= \left< g_{1},Tf \right>.
\end{equation}
Hence no boundary condition on $g$ is necessary. Consequently $T^{\ast}$ is given by
\begin{equation}\begin{aligned}
D(T^{\ast})&=\left\{g\in L^2(0,1):g'\in L^2(0,1)\right\}\\
T^{\ast}g&=-ig'.
\end{aligned}\end{equation}

The momentum operator given by Eq.(\ref{momentum}) is not self-adjoint. This happens because the chosen domain is so small, i.e., the restrictions on functions are so strong that allows the domain of $T^{\ast}$ to be extremely large.

Let us use Theorem \ref{theorem2} to find how many self-adjoint extensions does $T$ have.
\begin{equation}
(-ig'-\mp ig)=0\Rightarrow'g'=\mp g\Rightarrow g=e^{\mp x}.
\end{equation}
Therefore, there is one solution in $L^2(0,1)$ to each equation in (\ref{criterion}) and the self-adjoint extensions of $T$ are parametrized by the parameter $\theta$.

Now, the isometries from from $\ker(T^{\ast}-i)$ to $\ker(T^{\ast}+i)$ are given by $U:e^{-x} \to e^{i\gamma-1}e^{x}$, $\gamma\in\mathbb{R}$, because  
\begin{equation}
||e^{i\gamma-1}e^{x}||=\sqrt{\int^{1}_{0}{|e^{i\gamma-1}e^{x}|^{2}}dx}=\frac{1}{\sqrt{2}e}\sqrt{e^2-1}=||e^{-x}||.
\end{equation}
The extended domains of $T$ are given by
\begin{equation}
D^{\gamma}=\left\{f+e^{-x}+e^{i\gamma-1}e^{x}:f\in D(T)\right\}.
\end{equation}
Therefore, for a function $\psi\in D^{\gamma}$ we have
\begin{equation}\begin{aligned}
&\psi(0)=f(0)+1+e^{i\gamma-1}=1+e^{i\gamma-1}\\
&\psi(1)=f(1)+e^{-1}+e^{i\gamma}=e^{-1}+e^{i\gamma}.
\end{aligned}\end{equation}
Hence
\begin{equation}
\psi(1)=\frac{e^{-1}+e^{i\gamma}}{1+e^{i\gamma-1}}\psi(0)
\end{equation}
and, since $\left|\frac{e^{-1}+e^{i\gamma}}{1+e^{i\gamma-1}}\right|=1$, we have 
\begin{equation}
\psi(1)=e^{i\theta}\psi(0),\;\;\;\;\;\theta\in\mathbb{R}.
\end{equation}

The self-adjoint extensions of $T$ are given by
\begin{equation}\begin{aligned}
D(T_{\theta})&=\left\{\psi(x)\in L^2(0,1):\psi'\in L^2(0,1),\psi(1)=e^{i\theta}\psi(0)\right\}\\
T_{\theta}\psi&=-i\psi'.
\end{aligned}\end{equation}

This boundary condition assures conservation of probability.

In this example we saw that by relaxing the condition on functions belonging to $D(T)$ we could obtain a self-adjoint operator.
\end{exemplo}

\section{Quantum singularities}

In this section we follow the ideas introduced by Wald \cite{wald}, Horowitz and Marolf \cite{horowitz}. Just like in general relativity, where and extra information must be added at the singular points since we lose the aptness to predict the future of a particle following an incomplete world line, we define a spacetime as quantum mechanically singular if the time evolution of any wave packet is not completely determined by the initial wave data on a Cauchy surface.

Let $(M,g_{\mu\nu})$ be a static spacetime with a timelike Killing vector field $\xi^{\mu}$ and $t$ be the Killing parameter. 
 Klein-Gordon equation on this spacetime
\begin{equation}
\square\Psi=M^2\Psi 
\label{Klein-Gordon_0}
\end{equation}
can be splitted into a temporal and a spatial part, 
\begin{eqnarray}
\frac{\partial^2 \Psi}{\partial t^2}=-A\Psi=VD^{i}(VD_{i}\Psi)+M^2 V^2 \Psi,
\label{separada}
\end{eqnarray}
where $V^2=\xi^{\mu}\xi_{\mu}$ and $D_{i}$ is the spatial covariant derivative on a static spatial slice $\Sigma$ of the spacetime 
 (remember that the singular points are not part of the spacetime).

We may view $A$ as an operator on the Hilbert space $\mathcal{H}$ of the square-integrable functions on $\Sigma$ (following references 7, 12, 13 and 14). We chose this because we will consider an one-particle description of the field, which is mathematically equivalent to solve the first-order pseudo-differential equation
\begin{equation}
i\frac{\partial\Phi}{\partial t}=\sqrt{A}\Psi.
\label{square-root}
\end{equation}
We then take the physical axiom\cite{fulling} which says that the Hilbert space of possible quantum states of a single particle, in the theory governed by the relativistic wave equation (\ref{separada}), is the space of functions of the form
\begin{equation}
\Psi(t,x)=\int{\frac{d\sigma(j)}{\sqrt{2\omega_j}}\tilde{\phi}_{+}(j)\psi_j(x)e^{-i\omega_j t}},
\end{equation}
where $\tilde{\phi}_+\in L^2_\sigma$ and $\{\psi_j\}$ is a complete set of eingeinfunctions of the operator $A$ in Eq. (\ref{separada}). So, given $\tilde{\phi}_+\in L^2_\sigma$ we can define
\begin{equation}
\Psi_{NW}(t,x)\equiv\int{\tilde{\phi}_+(j)\psi_j(x)e^{-i\omega_j t}d\sigma(j)},
\end{equation}
where $NW$ means Newton-Wigner. Note that $\Psi_{NW}\neq\Psi$, but both represent the same state vector and solve Eq. (\ref{square-root}). We therefore have
\begin{equation}
\left\|\Phi_{NW}\right\|^2_\mu=\int{\left|\Psi_{NW}(t,x)\right|^2d\mu}=\left\|\tilde{\phi}_+\right\|^2_\sigma,
\end{equation}
which is independent of time (here $d\mu$ is the proper element on $\Sigma$). So it makes sense to intepret $\left|\Psi_{NW}(t,x)\right|^2$ as the probability density for observations of $x$ at time $t$. This justify our choice to take our Hilbert space as being $L^2$.

But of course the interpretation of $\left|\Psi_{NW}(t,x)\right|^2$ as the probability density has some problems as pointed out by Fulling\cite{fulling}. In particular, a particle localized in a compact set at time $t$ has nonzero probability of being detected at any given point at an instant later. This problem is solved by taking the negative frequence solutions into account. It makes clear that, despite the fact that our choice of the Hilbert space seems appropriate for an one-particle description, it is by no means as unique as in ordinary quantum mechanics. In fact, Ishibashi and Hosoya\cite{ishibashi} used the Sobolev space as the natural Hilbert space of the wave functions. The physical idea behind this choice is that they could prepare an initial data only with finite energy. They showed that in some cases, operators which are not essentially self-adjoint in $L^2$ are essentially self-adjoint if we use the Sobolev space instead.


To find the domain of the operator $A$, $D(A)$, is a more difficult task and 
generally no information is provided. So, a minimum domain is 
taken (a core), where the operator can be defined and which does not 
enclose the spacetime 
singularities. An appropriate set is $C_{0}^{\infty}(\Sigma)$, the set of the
 smooth function of compact support on $\Sigma$. But the chosen 
 domain is so small, i.e., the restrictions on functions are so strong, that the domain of the Hilbert adjoint operator $A^{\ast}$ 
 is extremely large and is composed of all functions $\psi$ in $L^2(\Sigma,V^{-1}d\mu)$ such that $A\psi \in L^2$. Then, $A$ 
 is not self-adjoint. Hence, we are faced with the problem to find self-adjoint extensions of $A$ and to discover if it has only 
 one or many of such extensions. On the Hilbert space
 $L^2(\Sigma, V^{-1}d\mu)$, where $d\mu$ is the proper element on $\Sigma$, it 
is not difficult to show (integrating by parts) that the operator 
$(A,C_{0}^{\infty}(\Sigma))$ is a positive symmetric operator. Then, 
self-adjoint extensions always exist \cite{reed} (at least
 the Friedrichs extension). 

If $A$ has only one self-adjoint extension (the closure $\overline{A}$ of $A$), then $A$ is essentially self-adjoint and, since we are worried with a one particle description, not a field theory, 
 the positive frequency solution satisfy
\begin{equation}
i\frac{\partial \Psi}{\partial t}=(\overline{A})^{1/2}\Psi,
\end{equation}  
and the evolution of a wave packet is uniquely determined by the 
initial data
\begin{equation}
\Psi(t,{\bf x})=e^{-it(\overline{A})^{1/2}}\Psi(0, {\bf x}).
\end{equation}
We say that the spacetime is quantum mechanically non-singular.

Now, if $A$ has many self-adjoint extensions $A_{\alpha}$, where $\alpha$ is a real parameter, we must choose one 
in order to evolve the wave packet. Any solution of the form
\begin{equation}
\Psi(t, {\bf x})=e^{-it(A_{\alpha})^{1/2}}\Psi(0, {\bf x}),
\end{equation} 
is a good solution and an extra information must be given to tell us which one has to be chosen. The spacetime 
 is said quantum mechanically singular and we can clearly see the resemblance to the classical case.
 
\section{Study of wave packets in classically singular spacetimes}

The first example of a classical singular theory, which becomes nonsingular in the view of quantum mechanics, is the nonrelativistic hydrogen atom. The imposition of quadratic integrability of the solutions of Schr\"odinger equation are sufficient to provide a complete set of eingenfunctions. Given an initial wave packet, its time evolution is uniquely determined.

Another example, now of a singularity which remains singular when tested by quantum mechanics, is the nonrelativistic
particle trapped in a 1-dimensional box. A boundary condition is necessary on both edges (it is usually taken
$\psi(0)=\psi(1)=0$) in order to evolve uniquely the wave packet.

Let us now study in a complete way some classical spacetimes already found in the literature.

\subsection{Global Monopole}

The metric around a global monopole \cite{barriola} is given by
\begin{equation}
ds^2=-dt^2+dr^2+\alpha^2r^2(d\theta^2+\sin^2{\theta}d\phi^2),
\label{metric}
\end{equation}
where $\alpha^2=(1-8\pi G\eta^2)$ and $\eta$ is the spontaneous symmetry breaking scale. It represents a symmetric cloud of cosmic strings, with all the strings forming the cloud intersecting at a single point $r = 0$ \cite{letelier}.

Klein-Gordon equation in this spacetime is given by \cite{monopole}
\begin{equation}
\frac{\partial^2\Psi}{\partial t^2}=\frac{1}{r^2}\frac{\partial}{\partial r}\left(r^2\frac{\partial \Psi}{\partial r} \right)+\frac{1}{\alpha^2r^2\sin{\theta}}\frac{\partial}{\partial \theta}\left(\sin{\theta}\frac{\partial \Psi}{\partial \theta}\right)+\frac{1}{\alpha^2r^2\sin^{2}{\theta}}\frac{\partial^2\Psi}{\partial \varphi^2}-M^2\Psi.
\label{klein monopole}
\end{equation}

By separating variables $\psi=R(r)Y_{l}^{m}(\theta,\varphi)$, Eq. (\ref{criterion}) reads
\begin{equation}
\frac{d^2R(r)}{dr^2}+\frac{2}{r}\frac{dR(r)}{dr}+\left[(\pm i-M^2)-\frac{l(l+1)}{\alpha^2r^2}\right]R(r).
\label{monopole equation}
\end{equation}

The solution of the above equation near infinity is given by
\begin{equation}   
R(r)=\frac{1}{r}[C_1e^{\beta r}+C_2e^{-\beta r}],
\label{solucaoinfinito}
\end{equation}
where 
\begin{equation}
\beta=\frac{1}{\sqrt{2}}\bigg[(\sqrt{1+M^4}+M^2)^{1/2} \mp 
i (\sqrt{1+M^4}-M^2)^{1/2}\bigg].
\end{equation}

This solution is square integrable only if $C_1=0$. So the asymptotic behavior of $R(r)$ is given by $R(r)\sim\frac{1}{r}e^{-\beta r}$. 

Near $r=0$, solution of Eq. (\ref{monopole equation}) is given by $r^{\gamma}$, with 
\begin{equation}
\gamma=\frac{-1\pm\sqrt{1+4\frac{l(l+1)}{\alpha^2}}}{2}.
\end{equation}

For $\gamma=-\frac{1}{2}+\frac{1}{2}\sqrt{1+4\frac{l(l+1)}{\alpha^2}}$ the solution $R(r)\sim r^{\gamma}$ is square-integrable near $r=0$. For $\gamma=-\frac{1}{2}-\frac{1}{2}\sqrt{1+\frac{l(l+1)}{\alpha^2}}$, $r^{\gamma}$ is square integrable only if $l=0$.  Therefore near origin we have
\begin{equation}
R_0(r)=\tilde{C}_1+\tilde{C}_{2}r^{-1}
\label{near0}
\end{equation} 
and we can adjust the constants in Eq. (\ref{near0}) to meet the asymptotic behavior $R_{0}(r)\sim \frac{1}{r}e^{-\beta r}$ \cite{extra}. There is one solution for each sign in Eq. (\ref{criterion}), so there is a one-parameter family of self-adjoint extensions of $A$. The spacetime is quantum mechanically singular.
 
The positive frequency solutions of Eq. (\ref{klein monopole}) are given by
\begin{equation}
\Psi(t,r,\theta,\varphi)=e^{-i\omega t}R_{\omega,l,m}Y_{l}^{m}(\theta, \varphi), 
\end{equation}
where
\begin{equation}
R_{\omega,l,m}(r)=A\frac{J_{\delta l}(kr)}{\sqrt{kr}}+B\frac{N_{\delta l}(kr)}{\sqrt{kr}}
\end{equation}
and
\begin{equation}
\delta_{l}=\frac{1}{2}\sqrt{1+4\frac{l(l+1)}{\alpha^2}}.
\end{equation}

The functions $J_{\delta_l}$ are always square-integrable near the origin and $N_{\delta_l}$ is square integrable only if $l=0$. For this value of $l$, the boundary condition on the function $R(r)$ is given by \cite{monopole,ishibashi}
\begin{equation}
R_{\omega,0,0}(r)=
\left\{\begin{aligned}
&\frac{\cos{kr}}{r}+\frac{1}{ak}\frac{\sin{kr}}{r}\;\;\;&a\neq0\\
&\frac{\cos{kr}}{r} & a=0.
\end{aligned}\right.
\end{equation}
so the general solution of Eq. (\ref{klein monopole}) is given by
\begin{equation}
\Psi_{a}=\int{\text{d}\omega e^{-i\omega t}\sum_{l=0}^{\infty}\sum_{m=-l}^{l}C(\omega,l,m)R_{\omega,l,m}(r)Y_{l}^{m}(\theta,\varphi)},
\end{equation}
with
\begin{equation}
R_{\omega,l,m}=\frac{J_{\delta_{l}}(kr)}{\sqrt{kr}}\;\;\; l\neq0.
\end{equation}

In order to choose one solution of Eq. (\ref{klein monopole}) we must fix one value of the parameter $a$. This choice is somewhat arbitrary since there is nothing in the theory which indicates the right choice.

\subsection{Cosmic string}

Klein-Gordon equation in the spacetime of a cosmic string with metric
\begin{equation}
ds^2=-dt^2+dr^2+\beta^2r^2d\phi^2+dz^2
\end{equation}
is given by \cite{helliwell2}
\begin{equation}
-\Phi_{,tt}+\Phi_{,rr}+\frac{1}{r}\Phi_{,r}+\frac{1}{\beta^2r^2}\Phi_{,\phi\phi}+\Phi_{,zz}=M^2\Phi.
\end{equation}

Equation (\ref{criterion}) now reads $(\nabla^2\pm i)\Phi=0$ where
\begin{equation}
\nabla^2=\frac{\partial^2}{\partial r^2}+\frac{1}{r}\frac{\partial}{\partial r}+\frac{1}{\beta^2 r^2}\frac{\partial^2}{\partial \phi^2}+\frac{\partial^2}{\partial z^2}.
\end{equation}

After separating variables in the form $\Phi=e^{im\phi}e^{ikz}R(r)$ we have
\begin{equation}
R''+\frac{1}{r}R'+\left[(\pm i-k^2)-\frac{m^2}{\beta^2r^2}\right]R=0.
\label{cosmic string}
\end{equation}

Near infinity the asymptotic solution is
\begin{equation}
R(r)=\frac{1}{\sqrt{r}}\left(C_1 e^{\alpha r}+C_2 e^{-\alpha r}\right)
\end{equation}
where
\begin{equation}
\alpha=\frac{1}{\sqrt{2}}\left[(\sqrt{1+k^4}+k^2)^{1/2}\mp i\left(\sqrt{1+k^4}-k^2\right)^{1/2}\right].
\end{equation}

This solution  is square integrable only if $C_1=0$. Near $r=0$ the solution of equation (\ref{cosmic string}) is given by $R(r)\sim r^{\pm\left|m/\beta\right|}$.  Both solutions are square integrable only if $\left|m/\beta\right|<1$ \cite{helliwell2}, i. e., $m=0$. Therefore there is a solution of equation (\ref{criterion}) and the spacetime is quantum mechanically singular.

For free spin-$1/2$ particles we have the Dirac equation
\begin{equation}
\left\{i\gamma^{(0)}\partial_t+i\gamma^{(r)}\left[\partial_r-\frac{1}{2r}\left(\frac{1-\beta}{\beta}\right)\right]+\frac{i}{\beta r}\gamma^{(\theta)}\left(\partial_\theta+\partial_t\right)+i\gamma^{(3)}\partial_z-m\right\}\Psi=0,
\label{dirac cosmic}
\end{equation} 
where $\gamma^{(r)}=\cos{\theta}\gamma^{(1)}+\sin{\theta}\gamma^{(2)}$ and $\gamma^{(\theta)}=-\sin{\theta}\gamma^{(1)}+\cos{\theta}\gamma^{(2)}$ and $\gamma^{(\mu)}$ are given in terms of the Pauli matrices by
\begin{equation}
\gamma^{(0)}=\begin{pmatrix} \sigma^{3}&0\\ 0&-\sigma^3\end{pmatrix}, \gamma^{(1)}=\begin{pmatrix} i\sigma^{2}&0\\ 0&-i\sigma^2\end{pmatrix},\gamma^{(2)}=\begin{pmatrix} -i\sigma^{1}&0\\ 0&i\sigma^1\end{pmatrix},\gamma^{(3)}=\begin{pmatrix} 0&1\\ -1&0\end{pmatrix}.
\end{equation}

Solution of equation (\ref{dirac cosmic}) is given by \cite{bezerra,shishkin}
\begin{equation}
\Psi(t,r,\phi,z)=\begin{pmatrix}\sqrt{E+m}R_1(r)\\i\sqrt{E+m}R_2(r)e^{i\phi}\end{pmatrix}e^{-iEt+im\phi+ikz}
\end{equation}
where
\begin{equation}
R_j^{''}+\frac{1}{r}R_{j}^{'}+\left[\left(E^2-m^2\right)-\frac{(\nu+j-1)^2}{r^2}\right]R_j=0\;\;\;\;\; (j=1,2)
\end{equation}
with $\nu=\frac{m+1/2}{\beta}-\frac{1}{2}$.

We must analyse the following equation
\begin{equation}
R_j^{''}+\frac{1}{r}R_{j}^{'}+\left[\left(E^2-m^2\right)-\frac{(\nu+j-1)^2}{r^2}\pm i\right]R_j=0.
\end{equation}

Both solutions of the above equation are square integrable near $r=0$ if, and only if, $\left|\nu+j-1\right|<1$ \cite{helliwell2}. Therefore the range of modes for which there is a quantum singularity is given by
\begin{equation}
-\frac{3}{2}<\frac{m+1/2}{\beta}<\frac{3}{2}.
\end{equation}
The spacetime is quantum mechanically singular when tested by Dirac particles too.

\subsection{BTZ spacetime}

The metric for the spinless BTZ spacetime \cite{banados} is given by
\begin{equation}
ds^2=-V(r)^2dt^2+V(r)^{-2}dr^2+r^2d\theta^2,
\label{metric btz}
\end{equation}
with the usual ranges of cylindrical coordinates and $V(r)$ is given by
\begin{equation}
V(r)^2=-m+\frac{r^2}{l^2},
\end{equation}
where $m$ is the mass parameter.

For $-1<m\leq 0$, there appears a continuous sequence of naked singularities at the origin. Since we are worried with naked singularities, we will work with these values of $m$.

As $r\to\infty$ the metric (\ref{metric btz}) takes the form
\begin{equation}
ds^2\approx -\bigg(\frac{r^2}{l^2}\bigg)dt^2+\bigg(\frac{r^2}{l^2}
\bigg)^{-1}dr^2+r^2d\theta^2.
\end{equation}
so that after separating variables in the form $\psi=R(r)e^{in\theta}$ equation (\ref{criterion}) takes the form (for large $r$) \cite{btz}
\begin{equation}
R_{n}''+\frac{3}{r}R_{n}'=0,
\end{equation}
whose solution is
\begin{equation}
R_{n}(r)=C_{1n}+C_{2n}r^{-2},
\end{equation}
where $C_{1n}$ and $C_{2n}$ are arbitrary constants. It is clear that $R(r)\in L^2$ if, and only if $C_1n=0$. To analyse the case $r\to 0$ we note that, after a redefinition of the coordinates ($t\to\alpha t,r\to\alpha^{-1}r$) \cite{btz} the metric (\ref{metric btz}) takes the form
\begin{equation}
ds^2\approx -dt^2+dr^2+\alpha^2r^2d\theta^2.
\end{equation} 

This represents a conical singularity. For this metric, the radial portion of equation (\ref{criterion}) is given by
\begin{equation}
R_{n}''+\frac{1}{r}R_{n}'+\bigg[\pm i-\frac{n^2}{\alpha^2r^2}\bigg]R_{n}=0.
\end{equation}

The general solution of the above equation is
\begin{equation}
R_{n}(r)=A_{n}J_{|n/\alpha|}(kr)+B_{n}N_{|n/\alpha|r},
\label{solution}
\end{equation}
with $k=\sqrt{i}$. Bessel function is always square integrable near the origin but $N_{\left|n/\alpha\right|}(r)$ is square integrable only for $n=0$. Then, for $n=0$ there is a solution of equation (\ref{criterion}) in $L^2$. The spacetime is quantum mechanically singular.

Because near $r=0$ the spacetime is similar to a conic spacetime, we can use the results of Ref. \cite{kay} where the boundary conditions necessary to turn self-adjoint the spatial portion of the wave operator in a conic spacetime where studied. They are  
\begin{equation}
 \begin{array}{ll}
\lim_{r\to 0}{\big\{\big[\ln(qr/2)+\gamma \big]rR_{0}'(r)-R_{0}\big\}=0},\; &q\in(0,\mu],\\ 
\lim_{r\to 0}{rR_{0}'(r)=0}, &q=0,
\end{array} 
\label{cond}
\end{equation}
where $\mu$ is the particle mass and $\gamma$ is Euler-Mascheroni constant. Note that if we are working with massless particles $\mu=0$ we must choose the boundary condition
\begin{equation}
\lim_{r\to 0}{rR_{0}'(r)=0}
\end{equation} 
so that the spacetime is nonsingular when tested by massless particles.

\subsection{Spherical Spacetimes}

Consider the spherical spacetimes with metric \cite{konkowski2}
\begin{equation}
ds^2=-dt^2+dr^2+r^{2p}(d\theta^2+\sin^2{\theta}d\phi^2)
\end{equation}

After separating variables in the form $\psi\sim f(r)Y(\text{angles})$ equation (\ref{criterion}) becomes
\begin{equation}
f''+\frac{2p}{r}f'+\left[\pm i-\frac{c}{r^{2p}}\right]f=0
\label{eq}
\end{equation}
where $c$ is a constant related with the angular variables. Let us concentrate in solutions with spherical symmetry, i.e., $c=0$. 

Near $r=0$ equation (\ref{eq}) reads
\begin{equation}
f''+\frac{2p}{r}f'=0.
\end{equation}
One solution goes like a constant while the other goes like $r^{1-2p}$. A constant is clearly square integrable near the origin, but $r^{1-2p}$ is square integrable only if $p<3/2$. Therefore the spacetime is quantum mechanically singular for $p<3/2$ and quantum mechanically regular for $p\geq3/2$ when tested by spherical symmetric test particles.

\section{Concluding Remarks}

A review of quantum singularities in static spacetimes following the ideas of Horowitz and Marolf was presented. A brief summary of the theory of unbounded operators in Hilbert spaces was given and the main results about essentially self-adjoint operators was established. We also studied quantum test particles in many classically singular spacetimes and found that there it is easier to find spacetimes where the introduction of quantum mechanics does not exclude the singularities. These examples were found in the literature and in particular we saw a curious result in the BTZ spacetime. There, the singularity remains when tested by massive particles but is excluded when tested by massless ones. Despite the fact that the theory of Horowitz and Marolf does not exclude the singularities in all classically singular spacetimes, it gives us evidence that a full quantum theory of gravity may remove the singular points of the classical theory.

\acknowledgements 
We thank Fapesp for financial  support and  P.S.L. also thanks  CNPq.



\begin{thebibliography}{99}
\bibitem{extra}
We say that $r=\infty$ is in the point-limit case and $r=0$ is in the circle-limit case. When this happens, we can assure that there is one solution of Eq. (\ref{criterion}). See Refs. \cite{reed} and \cite{richtmyer}.
\end{thebibliography}
\end{document}